\documentclass[amsmath,amssymb,aps,prd,reprint,superscriptaddress]{revtex4-1}

\usepackage{graphicx}
\usepackage{dcolumn}
\usepackage{bm}
\usepackage{comment}
\usepackage{here}

\begin{document}


\title{Demonstration of length control for a filter cavity with coherent control sidebands}


\author{Naoki Aritomi}
\email[]{naritomi@caltech.edu}
\affiliation{National Astronomical Observatory of Japan (NAOJ), Mitaka City, Tokyo 181-8588, Japan}
\affiliation{LIGO Hanford Observatory, Richland, Washington 99352, USA}

\author{Yuhang Zhao}
\affiliation{Institute for Cosmic Ray Research (ICRR), KAGRA Observatory, The University of Tokyo, Kashiwa City, Chiba 277-8582, Japan}

\author{Eleonora Capocasa}
\affiliation{Universit\'e de Paris, CNRS, Astroparticule et Cosmologie, F-75013 Paris, France}

\author{Matteo Leonardi}
\affiliation{National Astronomical Observatory of Japan (NAOJ), Mitaka City, Tokyo 181-8588, Japan}

\author{Marc Eisenmann}
\affiliation{National Astronomical Observatory of Japan (NAOJ), Mitaka City, Tokyo 181-8588, Japan}

\author{Michael Page}
\affiliation{National Astronomical Observatory of Japan (NAOJ), Mitaka City, Tokyo 181-8588, Japan}

\author{Yuefan Guo}
\affiliation{Nikhef, Science Park 105, 1098 XG Amsterdam, Netherlands}

\author{Eleonora Polini}
\affiliation{Laboratoire d'Annecy de Physique des Particules (LAPP),\\Universit\'e Savoie Mont Blanc, CNRS/IN2P3, F-74941 Annecy, France}

\author{Akihiro Tomura}
\affiliation{The University of Electro-Communications, Chofu City, Tokyo 182-8585, Japan}

\author{Koji Arai}
\affiliation{LIGO Laboratory, California Institute of Technology, Pasadena, California 91125, USA}

\author{Yoichi Aso}
\affiliation{National Astronomical Observatory of Japan (NAOJ), Mitaka City, Tokyo 181-8588, Japan}

\author{Martin van Beuzekom}
\affiliation{Nikhef, Science Park 105, 1098 XG Amsterdam, Netherlands}

\author{Yao-Chin Huang}
\affiliation{National Tsing Hua University, Hsinchu City, 30013, Taiwan}

\author{Ray-Kuang Lee}
\affiliation{National Tsing Hua University, Hsinchu City, 30013, Taiwan}

\author{Harald L\"{u}ck}
\affiliation{Max Planck Institute for Gravitational Physics (Albert Einstein Institute), D-30167 Hannover, Germany}

\author{Osamu Miyakawa}
\affiliation{Institute for Cosmic Ray Research (ICRR), KAGRA Observatory, The University of Tokyo, Hida City, Gifu 506-1205, Japan}

\author{Pierre Prat}
\affiliation{Universit\'e de Paris, CNRS, Astroparticule et Cosmologie, F-75013 Paris, France}

\author{Ayaka Shoda}
\affiliation{National Astronomical Observatory of Japan (NAOJ), Mitaka City, Tokyo 181-8588, Japan}

\author{Matteo Tacca}
\affiliation{Nikhef, Science Park 105, 1098 XG Amsterdam, Netherlands}

\author{Ryutaro Takahashi}
\affiliation{National Astronomical Observatory of Japan (NAOJ), Mitaka City, Tokyo 181-8588, Japan}

\author{Henning Vahlbruch}
\affiliation{Max Planck Institute for Gravitational Physics (Albert Einstein Institute), D-30167 Hannover, Germany}

\author{Marco Vardaro}
\affiliation{Nikhef, Science Park 105, 1098 XG Amsterdam, Netherlands}

\author{Chien-Ming Wu}
\affiliation{National Tsing Hua University, Hsinchu City, 30013, Taiwan}

\author{Matteo Barsuglia}
\affiliation{Universit\'e de Paris, CNRS, Astroparticule et Cosmologie, F-75013 Paris, France}

\author{Raffaele Flaminio}
\affiliation{Laboratoire d'Annecy de Physique des Particules (LAPP),\\Universit\'e Savoie Mont Blanc, CNRS/IN2P3, F-74941 Annecy, France}


\date{\today}

\begin{abstract}
For broadband quantum noise reduction of gravitational-wave detectors, a frequency-dependent squeezed vacuum field realized using a filter cavity is the most promising technique and will be implemented in Advanced LIGO and Advanced Virgo in the fourth observing run. To obtain the benefit of frequency-dependent squeezing, the length and alignment of the filter cavity with respect to the squeezed vacuum field must be accurately controlled. To this purpose, a new length and alignment control scheme for a filter cavity, using coherent control sidebands, was suggested [Phys. Rev. D \textbf{102}, 042003 (2020)]. The coherent control sidebands are already used to control the squeezing angle in squeezed vacuum sources for gravitational-wave detectors. As both the squeezed vacuum field and coherent control sidebands have the same mode-matching conditions and almost the same frequency, the length and alignment of the filter cavity with respect to the squeezed vacuum field can be accurately controlled with this scheme. In this paper, we experimentally demonstrate the new control scheme for a filter cavity with coherent control sidebands. In addition to the conventional filter cavity control with the green field, we succeed in controlling the length of a 300-m filter cavity with coherent control sidebands and reduce the filter cavity length noise (rms) from 6.8 to 2.1 pm.
\end{abstract}


\maketitle

\section{Introduction\label{introduction}}
Gravitational waves (GWs) were detected for the first time by Advanced LIGO in 2015 \cite{PhysRevLett.116.061102}, and since then many more GWs have been observed by Advanced LIGO and Advanced Virgo \cite{PhysRevX.9.031040, PhysRevX.11.021053, GWTC-3}. To increase the number of detections, the sensitivity of the detectors must be constantly improved. One of the main noise sources for GW detectors is quantum noise. Quantum noise is divided into shot noise (which limits the sensitivity at high frequencies) and radiation pressure noise (which limits the sensitivity at low frequencies). An effective way to reduce quantum noise is to inject a squeezed vacuum field into the interferometer \cite{PhysRevD.23.1693}. The reduction of quantum noise with squeezing was first realized at GEO600 \cite{NatPhys.7.12.962}, and it was implemented in Advanced LIGO and Advanced Virgo during the third observing run \cite{PhysRevLett.123.231107, PhysRevLett.123.231108}. However, a conventional frequency-independent phase-squeezed vacuum field increases radiation pressure noise at low frequencies, while it reduces shot noise at high frequencies \cite{Nature.583, PhysRevLett.125.131101}. For broadband quantum noise reduction, a frequency-dependent squeezed vacuum field produced with a filter cavity is the technique being implemented in GW detectors \cite{PhysRevD.65.022002}. Advanced LIGO and Advanced Virgo plan to implement frequency-dependent squeezing with 300 m filter cavities in the fourth observing run \cite{scenario}. In order to achieve frequency dependence below 100 Hz, which is necessary for broadband quantum noise reduction in GW detectors, the filter cavity has to be operated in a detuned configuration (i.e., off resonance of the carrier) and it needs a storage time of about 3 ms. Frequency-dependent squeezing below 100 Hz was recently demonstrated \cite{PhysRevLett.124.171101, PhysRevLett.124.171102}. 
\par One of the main challenges in the production of frequency-dependent squeezing using filter cavities is the length and alignment control of the filter cavity itself. As squeezing is a vacuum state with no coherent amplitude, it is not suitable to provide the error signals necessary for the control, and therefore, auxiliary fields are needed. In previous experiments \cite{PhysRevLett.124.171101, PhysRevD.105.082003} the filter cavity was controlled with an auxiliary green field with a wavelength of 532 nm, while the squeezed field is at the GW detector laser wavelength, 1064 nm. However, the length and alignment control of the filter cavity with the green field does not ensure the alignment of the squeezed field to the filter cavity because the relative alignment of the green and squeezed fields can drift. In addition, fluctuation of the relative phase delay between the green and squeezed fields induced by anisotropies of the cavity mirror coating leads to a detuning fluctuation \cite{PhysRevD.105.082003}.
\par To solve this problem, a new length and alignment control scheme for the filter cavity, whose error signal is provided by coherent control sidebands (CCSBs), was suggested (we refer to this scheme as CCFC in this paper) \cite{PhysRevD.102.042003}. The CCSBs are included in all of the squeezed vacuum sources for GW detectors, and they are used to control the squeezing angle \cite{PhysRevA.75.043814}. As the CCSBs are produced inside an optical parametric oscillator (OPO) together with the squeezed vacuum field, they have the same mode-matching conditions and almost the same frequency. The relative frequency of the carrier and CCSBs can be accurately controlled with a frequency-offset phase-locked loop and can be tuned so that the carrier is properly detuned. Such a frequency difference is only a few MHz, and any possible phase shifts due to the mirror coating are negligible. Therefore, length and alignment control with CCSBs ensures proper length and alignment of the filter cavity with respect to the squeezed vacuum field.
\par For the length control of the filter cavity, a scheme that uses an additional auxiliary field called a resonant locking field (RLF) was successfully tested in a recent work \cite{PhysRevLett.124.171102}. The RLF is injected into the OPO and copropagates with the CCSBs. In the RLF scheme, the RLF resonates inside the filter cavity and gives the filter cavity length signal, while the CCSBs do not. Compared with the RLF scheme, the CCFC scheme has the advantage that it does not require any additional auxiliary field.
\par In this paper we experimentally demonstrate the new control scheme for a filter cavity with coherent control sidebands. In addition to the conventional filter cavity control with the green field, we succeed in controlling the length of a 300 m filter cavity with CCFC and reduce the filter cavity length noise (rms) from 6.8 to 2.1 pm.
\par This paper is organized as follows. In Sec. \ref{Principle} the principle of the filter cavity control scheme using CCSBs is reviewed. In Sec. \ref{Experimental setup} the experimental setup for the demonstration of CCFC scheme is explained. In Sec. \ref{Results} the experimental results of the CCFC error signal, filter cavity locking precision, and frequency-dependent squeezing with CCFC are presented.

\section{Principle} \label{Principle}
In this section, based on Ref. \cite{PhysRevD.102.042003}, we review the principle of the filter cavity control scheme using CCSBs. 

The CCSBs reflected by the filter cavity can be written as \cite{PhysRevD.102.042003}
\begin{eqnarray}
E_\mathrm{cc} &=& a_+ r_+ e^{i(\omega_0+\Omega_\mathrm{cc})t+i\phi_\mathrm{cc}} \nonumber\\
&+& a_- r_- e^{i(\omega_0-\Omega_\mathrm{cc})t + i(\phi_\mathrm{cc}+\phi_\mathrm{pump})},
\end{eqnarray}
where $\omega_0$ is the carrier frequency, $\Omega_\mathrm{cc}$ is the detuning of the coherent control field with respect to the carrier, $\phi_\mathrm{cc}$ is the phase of the coherent control field, and $\phi_\mathrm{pump}$ is the phase of the pump field. In the CCFC scheme, $\Omega_\mathrm{cc}$ is chosen as 
\begin{eqnarray}
\Omega_\mathrm{cc} = n\times \omega_\mathrm{FSR} + \Delta \omega_{\mathrm{fc},0}, \label{Omega_cc}
\end{eqnarray}
where $n$ is an integer number, $\omega_\mathrm{FSR} = 2\pi f_\mathrm{FSR}=\pi c/L_\mathrm{fc}$ is the free spectral range of the filter cavity, $L_\mathrm{fc}$ is the filter cavity length, and $\Delta \omega_{\mathrm{fc},0}$ is the optimal filter cavity detuning with respect to the carrier, which is defined as in Eq. (11) of Ref. \cite{PhysRevD.93.082004}. \\ 
$a_\pm$ is the amplitude of CCSBs and can be written as
\begin{equation}
a_+ = a_\mathrm{cc} \dfrac{1}{(1-x^2)},\ a_- = a_\mathrm{cc} \dfrac{x}{(1-x^2)},
\end{equation}
where $a_\mathrm{cc}$ is the amplitude of the coherent control field without the pump field, and $x = 1 - 1/\sqrt{g}$ is the OPO nonlinear factor, where $g$ is the nonlinear gain. \\
$r_\pm (\Delta \omega_\mathrm{fc}) = r_\mathrm{fc}(\pm \Delta \omega_{\mathrm{fc},0}, \Delta\omega_\mathrm{fc})$ is the complex reflectivity of the filter cavity and can be written as \cite{PhysRevD.90.062006}
\begin{eqnarray}
r_\mathrm{fc}(\pm \Delta \omega_{\mathrm{fc},0}, \Delta\omega_\mathrm{fc})  \simeq 1 - \dfrac{2-\epsilon}{1+i\xi(\pm \Delta \omega_{\mathrm{fc},0},\Delta\omega_\mathrm{fc})}, \label{r_fc}
\end{eqnarray}
where
\begin{eqnarray}
\epsilon &=& \dfrac{f_\mathrm{FSR}}{\gamma_\mathrm{fc}}\Lambda^2_\mathrm{rt}, \\
\xi (\pm \Delta \omega_{\mathrm{fc},0},\Delta\omega_\mathrm{fc}) &=& \dfrac{\pm \Delta \omega_{\mathrm{fc},0} 
-\Delta\omega_\mathrm{fc}}{\gamma_\mathrm{fc}}.
\end{eqnarray}
$\gamma_\mathrm{fc}$ is the filter cavity half-bandwidth, and $\Lambda_\mathrm{rt}^2$ are the filter cavity round trip losses. $\Delta\omega_\mathrm{fc}$ is a variable that represents the actual filter cavity detuning. 
\par The amplitude and phase of the filter cavity reflectivity for CCSBs can be written as
\begin{eqnarray}
\rho_{\pm}(\Delta\omega_\mathrm{fc})  &=& |r_\mathrm{fc}(\pm \Delta \omega_{\mathrm{fc},0},\Delta\omega_\mathrm{fc})| \nonumber\\
&=& \sqrt{1-\dfrac{(2-\epsilon)\epsilon}{1+\xi^2(\pm \Delta \omega_{\mathrm{fc},0},\Delta\omega_\mathrm{fc})}}, \\
\alpha_{\pm}(\Delta\omega_\mathrm{fc}) &=& \mathrm{arg}\{r_\mathrm{fc}(\pm \Delta \omega_{\mathrm{fc},0},\Delta\omega_\mathrm{fc})\}\nonumber\\
&=& \mathrm{arg}\{-1+\epsilon+\xi^2 (\pm \Delta \omega_{\mathrm{fc},0},\Delta\omega_\mathrm{fc}) \nonumber\\
&& +\ i(2-\epsilon)\xi(\pm \Delta \omega_{\mathrm{fc},0},\Delta\omega_\mathrm{fc})\}. \label{FC phase}
\end{eqnarray}

The filter cavity length signal can be obtained by detecting the beat note of the CCSBs:
\begin{eqnarray}
P_\mathrm{cc} &=& \left| a_+ r_+ e^{i(\omega_0 + \Omega_\mathrm{cc})t} + a_- r_- e^{i(\omega_0 - \Omega_\mathrm{cc})t + i\phi_\mathrm{pump}}\right|^2 \nonumber \\
&=& (\mathrm{DC\ term})\ +\ 2a_+ a_- \mathrm{Re}\{ r_+ r_-^\ast e^{i(2\Omega_\mathrm{cc}t-\phi_\mathrm{pump})}\}. \nonumber\\ \label{beat note}
\end{eqnarray}
Demodulating this signal by $\sin{(2\Omega_\mathrm{cc} t -\phi_\mathrm{dm,CCFC}})$ and low passing it, the CCFC error signal as a function of the filter cavity detuning $\Delta\omega_\mathrm{fc}$ is
\begin{eqnarray}
P_\mathrm{CCFC} &=& - a_+ a_- \rho_+ (\Delta\omega_\mathrm{fc}) \rho_- (\Delta\omega_\mathrm{fc}) \nonumber\\
&\times& \sin{(\alpha_+ (\Delta\omega_\mathrm{fc}) - \alpha_- (\Delta\omega_\mathrm{fc}) 
+  \phi_\mathrm{dm,CCFC} -  \phi_\mathrm{pump})}. \nonumber \\ \label{CCFC I phase} 
\end{eqnarray}
In general, the CCFC demodulation phase $\phi_\mathrm{dm,CCFC}$ can be written as
\begin{eqnarray}
\phi_\mathrm{dm,CCFC} = \phi_\mathrm{pump} + \alpha_- (\Delta\omega_{\mathrm{fc},0})+\pi + \delta \phi_\mathrm{dm, CCFC}, \nonumber \\ \label{CCFC demodulation phase}
\end{eqnarray}
where $\delta \phi_\mathrm{dm, CCFC}$ is an arbitrary phase.
\par
\begin{figure}[t]
\begin{center}
\includegraphics[width=15cm,bb= 100 0 3070 1100]{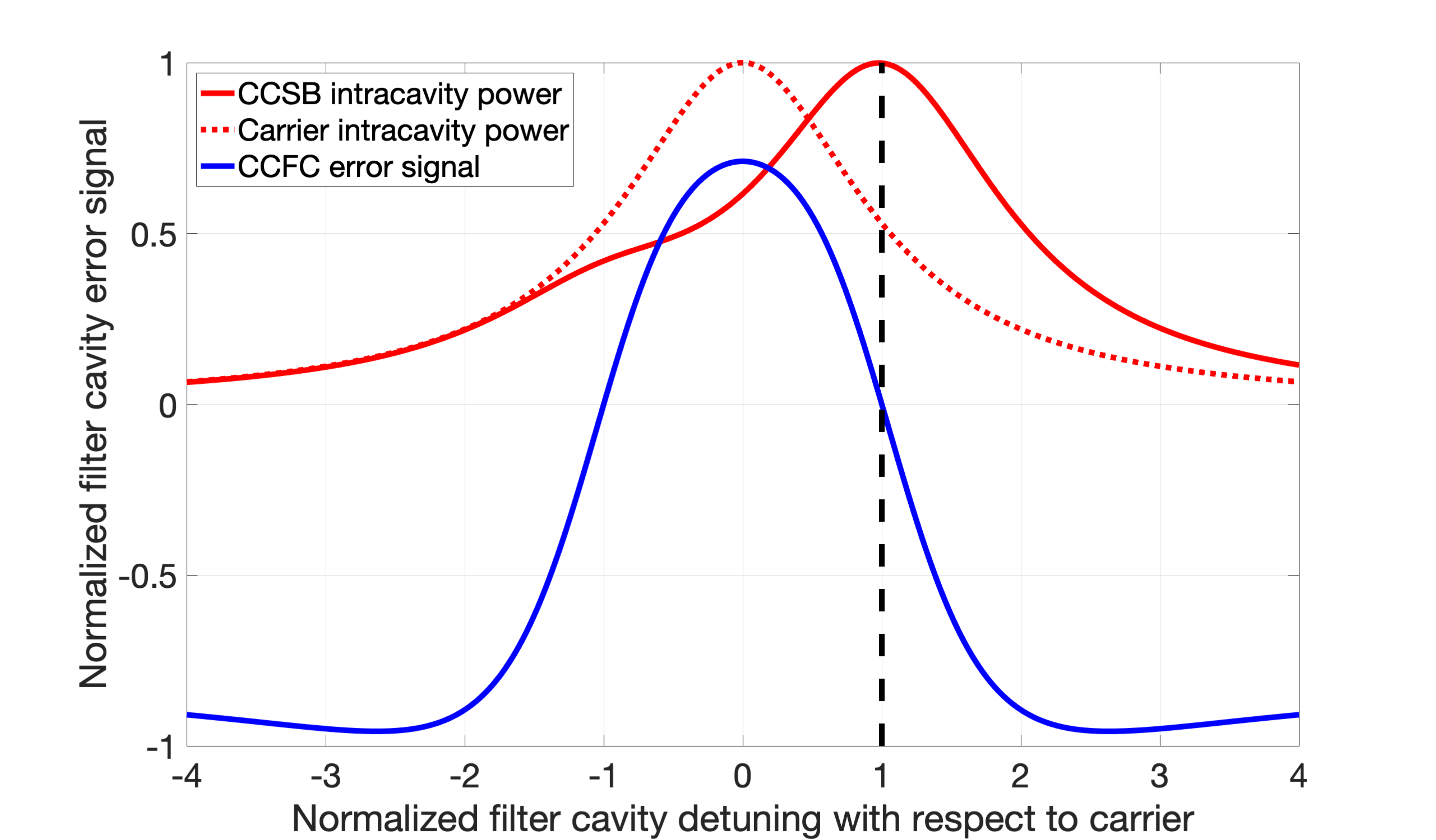}
\caption{CCFC error signal normalized with respect to $a_+a_-$. The horizontal axis is the filter cavity detuning normalized with respect to the optimal detuning $\Delta \omega_\mathrm{fc,0}$. The red solid and dashed lines are the intracavity power of CCSBs and carrier in the filter cavity normalized with respect to their maximum intracavity power, respectively. The blue line is the CCFC error signal (in phase). The CCFC error signal (in phase) becomes 0 when $\Delta \omega_\mathrm{fc}=\Delta \omega_\mathrm{fc,0}$.}
\label{CCFC theory}
\end{center}
\end{figure}

 In Fig. \ref{CCFC theory} we show the CCFC error signal (\ref{CCFC I phase}) normalized with respect to $a_+ a_-$ when $\delta \phi_\mathrm{dm, CCFC} = 0$. The parameters used in this calculation are shown in Table \ref{table1}.

\begin{table}[t]
\caption{\label{table1}%
Target parameters of a 300 m filter cavity for KAGRA \cite{PhysRevD.93.082004}.
}
\begin{ruledtabular}
\begin{tabular}{lcc}
\textrm{Parameter}&
\textrm{Symbol}&
\textrm{Value}\\
\colrule
Filter cavity length & $L_\mathrm{fc}$ & 300 m\\
Filter cavity half-bandwidth& $\gamma_\mathrm{fc}$ & $2\pi\times57.3$ Hz\\
Filter cavity detuning & $\Delta\omega_{\mathrm{fc},0}$ & $2\pi\times54$ Hz \\
Filter cavity finesse & $\mathcal{F}$ & 4360\\
Filter cavity input mirror transmissivity & $t_\mathrm{in}^2$ & 0.00136\\
Filter cavity round-trip losses & $\Lambda^2_\mathrm{rt}$ & 80 ppm\\
Propagation losses & $\Lambda^2_\mathrm{prop}$ & 10$\%$\\
Mode-mismatch losses & $\Lambda^2_\mathrm{mmFC}$ & 2$\%$\\
(squeezer-filter cavity) &&\\
Mode-mismatch losses & $\Lambda^2_\mathrm{mmLO}$ & 5$\%$\\
(squeezer-local oscillator) &&\\
Frequency-independent phase  &$\delta\phi$& 30 mrad\\
noise (rms) &&\\
Filter cavity length noise (rms)  &$\delta L_\mathrm{fc}$& 1 pm\\
Generated squeezing &$\sigma_\mathrm{dB}$& 9 dB\\
Nonlinear gain &$g$& 3.6\\
\end{tabular}
\end{ruledtabular}
\end{table}

\par The mode mismatch between the squeezed vacuum field (or CCSBs) and the filter cavity affects the CCFC error signal. In this paper, the mode mismatch means all of the higher-order modes including the misalignment. We express the squeezed vacuum field on the basis of the filter cavity mode,
\begin{eqnarray}
U_\mathrm{sqz} = \sum _{n=0} ^\infty a_n U_n,
\end{eqnarray}
where $U_n$ is the orthogonal basis of spatial modes and $U_0$ is the filter cavity fundamental mode. $a_n$ are complex coefficients and satisfy $\sum _{n=0} ^\infty |a_n|^2 = 1$.
\par
\begin{figure}[t]
\begin{center}
\includegraphics[width=15cm,bb= 100 0 3070 1100]{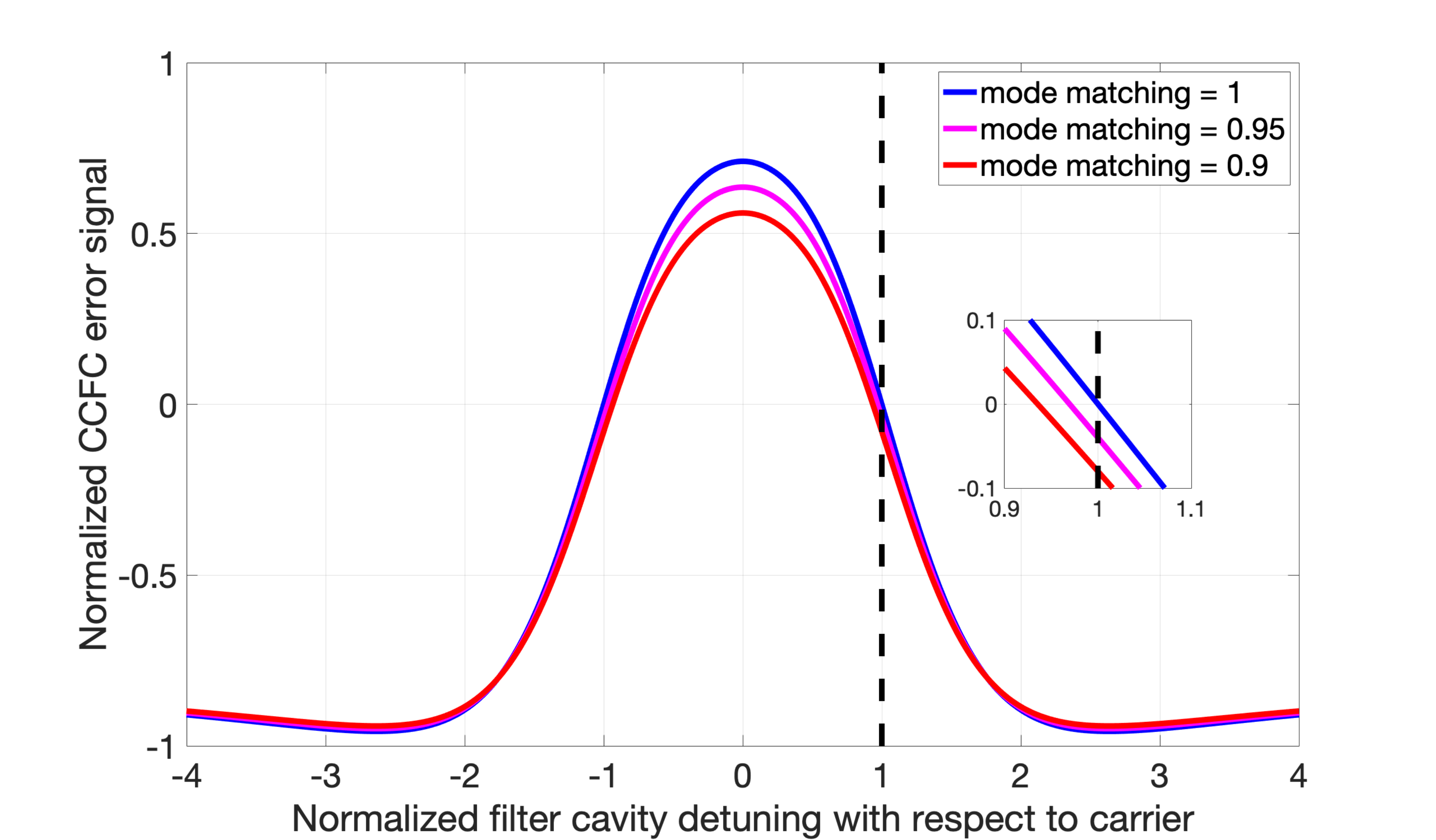}
\caption{CCFC error signal normalized with respect to $a_+a_-$ for different mode matching between the squeezed field and the filter cavity $|a_0|^2$. The horizontal axis is the filter cavity detuning normalized with respect to the optimal detuning $\Delta \omega_\mathrm{fc,0}$.}
\label{CCFC mismatch figure}
\end{center}
\end{figure}

 Considering the mode mismatch, the beat note of the CCSBs (\ref{beat note}) can be written as
\begin{eqnarray}
P_\mathrm{cc} &=& \left| a_+ \left(r_+ a_0 U_0 + \sum _{n=1} ^\infty a_n U_n\right) e^{i(\omega_0 + \Omega_\mathrm{cc})t} \right. \nonumber \\
&+& \left. a_- \left(r_-a_0 U_0 + \sum _{n=1} ^\infty a_n U_n \right) e^{i(\omega_0 - \Omega_\mathrm{cc})t + i\phi_\mathrm{pump}}\right|^2 \nonumber \\
&=& (\mathrm{DC\ term}) \nonumber \\
 &+& 2a_+ a_- \mathrm{Re}\left\{ (r_+ r_-^\ast |a_0|^2 + \sum _{n=1} ^\infty |a_n|^2)e^{i(2\Omega_\mathrm{cc}t-\phi_\mathrm{pump})}\right\}. \nonumber\\ \label{beat note with mode mismatch}
\end{eqnarray}
Here, we assume that all higher-order modes are far from the filter cavity resonance ($r_\pm \simeq 1$). Demodulating this signal by $\sin{(2\Omega_\mathrm{cc} t -\phi_\mathrm{dm,CCFC}})$ and low passing it, the CCFC error signal with the mode mismatch normalized with respect to $a_+ a_-$ can be written as
\begin{eqnarray}
 &&P_\mathrm{CCFC}/(a_+ a_-) \nonumber \\
 &=& |a_0|^2  \rho_+ \rho_- \sin{(\alpha_+  - \alpha_-  +  \alpha_{-,0} + \delta \phi_\mathrm{dm, CCFC})} \nonumber \\
 &+&(1-|a_0|^2) \sin{(\alpha_{-,0} + \delta \phi_\mathrm{dm, CCFC})}. \label{CCFC mismatch}
\end{eqnarray}
The CCFC error signal with the mode mismatch is shown in Fig. \ref{CCFC mismatch figure}. The zero-crossing point of the CCFC error signal changes due to the mode mismatch, and this causes the detuning change of the filter cavity. In Fig. \ref{CCFC mismatch figure}, the mode mismatch of 5 \% and 10 \% causes the detuning change of 1.6 and 3.5 Hz, respectively. If the mode mismatch is constant, the shift of the zero-crossing point caused by the mode mismatch is also constant and can be compensated by changing the frequency of the CC field or the demodulation phase of the CCFC error signal. Although the actual mode mismatch would fluctuate in time and it would be difficult to compensate it, the fluctuation of the zero-crossing point due to misalignment is expected to be reduced with alignment control of the filter cavity.

 \section{Experimental setup} \label{Experimental setup}
\begin{figure}[t]
\begin{center}
\includegraphics[width=10.5cm,bb= 0 0 3770 5250]{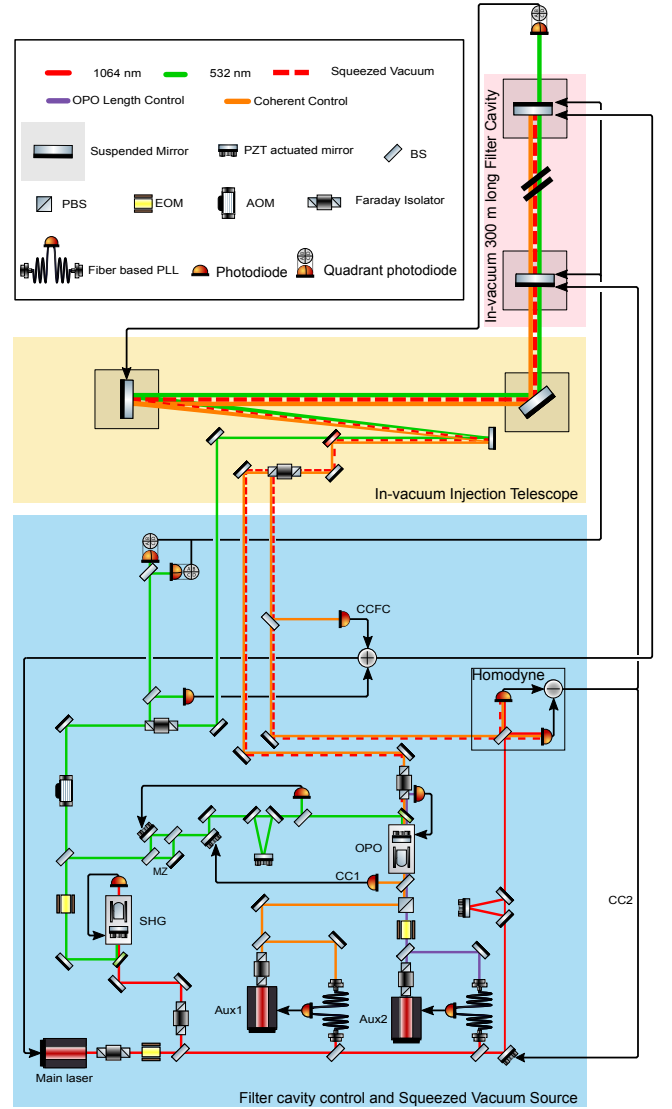}
\caption{Schematic of the experimental setup for CCFC. The PZT, BS, PBS, EOM, and PLL mean the piezoelectric, the beam splitter, the polarized beam splitter, the electro-optic modulator, and the phase-locked loop. A beam splitter with a transmissivity of 80 \% is placed in the infrared reflection path of the filter cavity to pick off the CCSBs. The CCSBs are detected with a resonant photodetector and demodulated at $2\Omega_\mathrm{cc} = 13.9940828$ MHz to obtain the CCFC error signal. The CCFC error signal is mixed with the error signal for the filter cavity control with the green field and fed back to the main laser frequency above 10 Hz and to the filter cavity end mirror below 10 Hz.}
\label{CCFC setup}
\end{center}
\end{figure}

The experimental setup for the demonstration of the CCFC is shown in Fig. \ref{CCFC setup}. The setup consists of an in-vacuum 300 m long filter cavity, an in-vacuum injection telescope, and an in-air squeezed vacuum source. The mirrors of the 300-m filter cavity and the two steering mirrors in the injection telescope are suspended by the vibration isolation system \cite{2002, doi:10.1063/1.1473225}.

\par The main laser is divided into the pump field for the second harmonic generator (SHG) and the local oscillator (LO) for the homodyne detector. The green field generated by the SHG is divided into the filter cavity control field and the pump field for the OPO. An acousto-optic modulator (AOM) is placed in the path of the green field for filter cavity control to adjust the relative frequency between the green and squeezed fields. A Mach-Zehnder interferometer (MZ) and a mode-cleaner cavity are placed in the path of the green field before the OPO to stabilize the power of the green field and clean the beam shape of the green field, respectively. 
\par There are two auxiliary lasers, which are phase locked to the main laser. Auxiliary laser 1 has the same polarization as the squeezed field and is used to control the squeezing angle (coherent control). Auxiliary laser 2 is polarized orthogonal to the squeezed field and is used for the OPO length control. The frequency difference between the main laser and auxiliary laser 1 for coherent control is chosen as $\Omega_\mathrm{cc} = 6.9970414$ MHz to satisfy Eq. (\ref{Omega_cc}), while that between the main laser and auxiliary laser 2 is defined by the coresonance condition of s and p polarizations inside the OPO, which depends on the crystal birefringence. 
\par To pick off the CCSBs, a beam splitter with transmissivity of 80 \% is placed in the infrared (IR) reflection path of the filter cavity at the expense of introducing an additional loss of 20 \%. In the real GW detectors, the CCSBs can be obtained at the reflection of an output mode cleaner without introducing the additional loss. The CCSBs are detected with a resonant photodetector and demodulated at $2\Omega_\mathrm{cc} = 13.9940828$ MHz to obtain the CCFC error signal. The CCFC error signal is mixed with the error signal for the filter cavity control with the green field and fed back to the main laser frequency above 10 Hz and to the filter cavity end mirror below 10 Hz. The crossover frequency of the CCFC and the filter cavity control with the green field is around 2 kHz: the CCFC is dominant below 2 kHz and the filter cavity control with the green field is dominant above 2 kHz.
\par For alignment control of the filter cavity, the input and end filter cavity mirrors are controlled by wavefront sensing with the green field. For beam pointing control, one of the suspended steering mirrors is controlled to fix the beam spot of the green field at the filter cavity transmission \cite{PhysRevD.105.082003}.
\par The coherent control consists of two control loops: the control loop to fix the relative phase between the pump field and the coherent control field (CC1), and the control loop to fix the relative phase between the coherent control field and the LO (CC2).
\par One of the noise sources for squeezing measurement is the backscattering noise. A part of the LO is leaked to the squeezing path, reflected by the filter cavity, and injected into the homodyne detector. The main scattering source is the motion of the suspended mirrors below 10 Hz. To reduce the backscattering noise, the CC2 control signal is sent to the phase shifter in the LO path above 10 Hz and to the filter cavity input mirror below 10 Hz.

\section{Results} \label{Results}
In this section, we present the experimental results of the CCFC error signal, filter cavity locking precision, and frequency-dependent squeezing with CCFC.

\subsection{CCFC error signal}
\begin{figure}[t]
\begin{center}
\includegraphics[width=15cm,bb= 100 0 3070 1100]{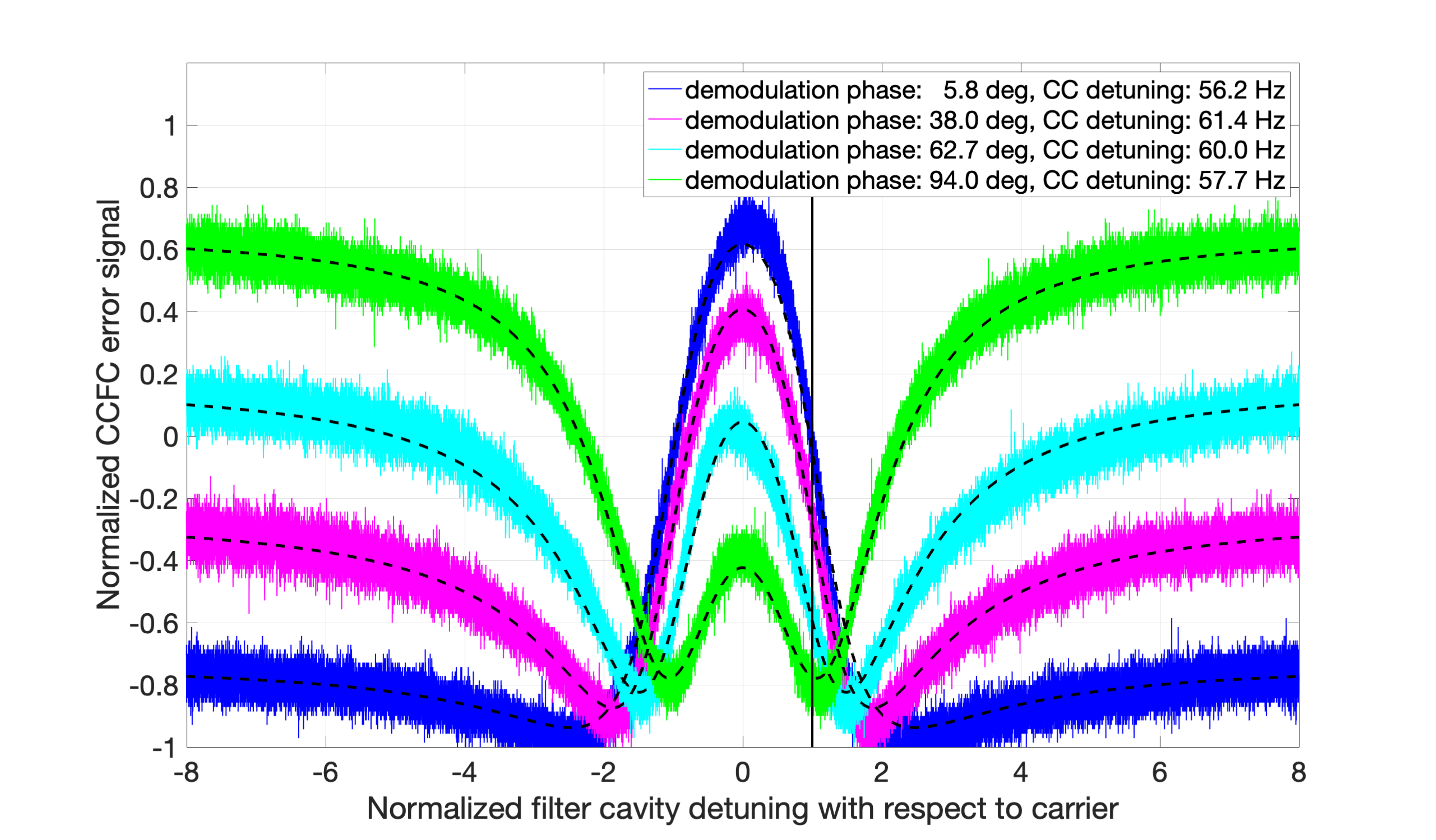}
\caption{Measured CCFC error signals normalized with respect to $a_+ a_-$ for different demodulation phases $\delta \phi_\mathrm{dm,CCFC}$. The horizontal axis is the filter cavity detuning normalized with respect to the optimal detuning of 54 Hz. The dashed black lines represent the
fitting of the CCFC error signals. The fitting parameters are the demodulation phase $\delta \phi_\mathrm{dm,CCFC}$, the CC detuning with respect to the carrier $\Delta \omega_\mathrm{fc,0}$, and the starting time of the scan. The mode matching between the CCSBs and the filter cavity is fixed to $|a_0|^2 = 0.94$, which is the measured value. The solid black line represents the optimal detuning.}
\label{CCFC error signal}
\end{center}
\end{figure}

The CCFC error signal is measured by locking the filter cavity with the green field and scanning the frequency applied to the AOM for the green field to scan the squeezing frequency with respect to the filter cavity resonance. The measured CCFC error signals with different demodulation phases are shown in Fig. \ref{CCFC error signal}. The measured data are fitted with the theoretical CCFC error signal (\ref{CCFC mismatch}). The fitting parameters are the demodulation phase $\delta \phi_\mathrm{dm, CCFC}$, the CC detuning with respect to the carrier $\Delta \omega_\mathrm{fc,0}$, and the starting time of the scan. The mode matching between the CCSBs and the filter cavity is fixed to $|a_0|^2 = 0.94$, which is the measured value.
The amplitude of the measured CCFC error signal is normalized with respect to $a_+ a_-$. This amplitude can be measured from the CCFC error signal when the CCSBs are off resonance of the filter cavity [$\alpha_+ (\Delta \omega_\mathrm{fc})= \alpha_- (\Delta \omega_\mathrm{fc}) = 0,\ \rho_+(\Delta \omega_\mathrm{fc}) = \rho_-(\Delta \omega_\mathrm{fc})=1$ in Eq. (\ref{CCFC I phase})] and the green pump phase $\phi_\mathrm{pump}$ is scanned.
\par The measured CCFC error signal is consistent with the theoretical one. The blue curve in Fig. \ref{CCFC error signal}, which is close to the in-phase CCFC error signal ($\delta \phi_\mathrm{dm,CCFC}=0$ deg), crosses zero around the optimal detuning at 54 Hz. Using this in-phase signal, we can lock the filter cavity around the optimal detuning as well as adjust the detuning by changing the relative frequency between the main laser and CC laser, or the CCFC demodulation phase. 

\subsection{Filter cavity locking precision with CCFC}
\begin{figure}[h] 
\begin{center}
\includegraphics[width=13.5cm,bb= 100 0 3100 1200]{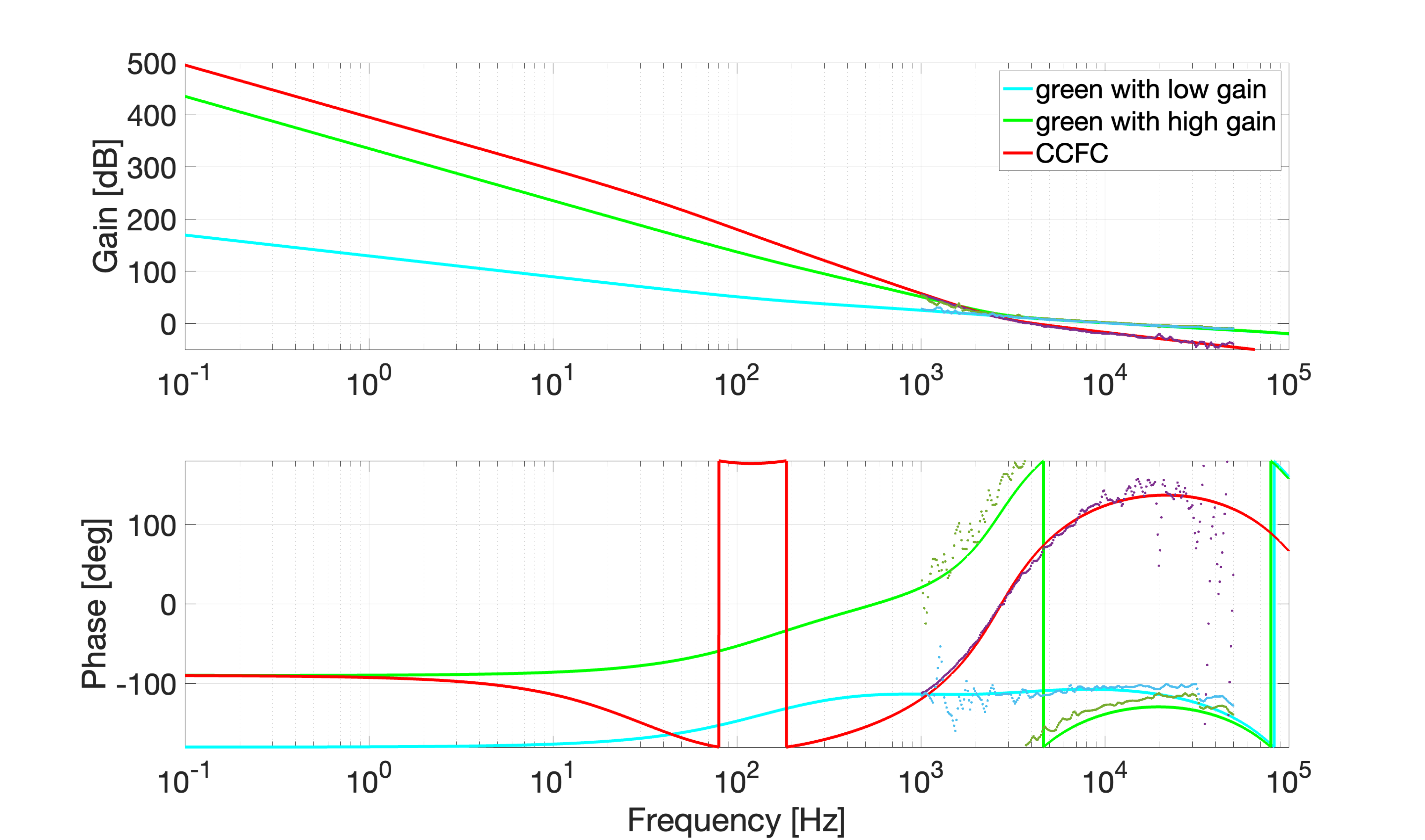}
\caption{Open-loop transfer functions of filter cavity control with different filters. The upper and lower panels show the gain and phase, respectively. The cyan and green curves are the open-loop transfer functions of the green control with low gain and high gain, respectively. The red curve is the open-loop transfer function of CCFC. The measured open-loop transfer functions of each control are shown in the frequency range of 1-50 kHz.}
\label{CCFC OLTF}
\end{center}
\end{figure}

To validate the effect of CCFC on the filter cavity locking precision, we measure the filter cavity locking precision with different control filters. The filter cavity locking precision can be obtained from the CCFC error signal. The filter cavity is controlled with three types of control filters. The first two filters, called ``high gain" and ``low gain", use the error signal of the green field and their open-loop transfer functions are shown in Fig. \ref{CCFC OLTF}. The two filters have almost the same unity gain frequency, but the high-gain filter has a significantly larger gain below the unity gain frequency. The third filter uses a combination of the CCFC error signal (below 1.6 kHz) and the green error signal (above 1.6 kHz). 

\begin{figure}[h] 
\begin{center}
\includegraphics[width=13.5cm,bb= 100 0 3100 1200]{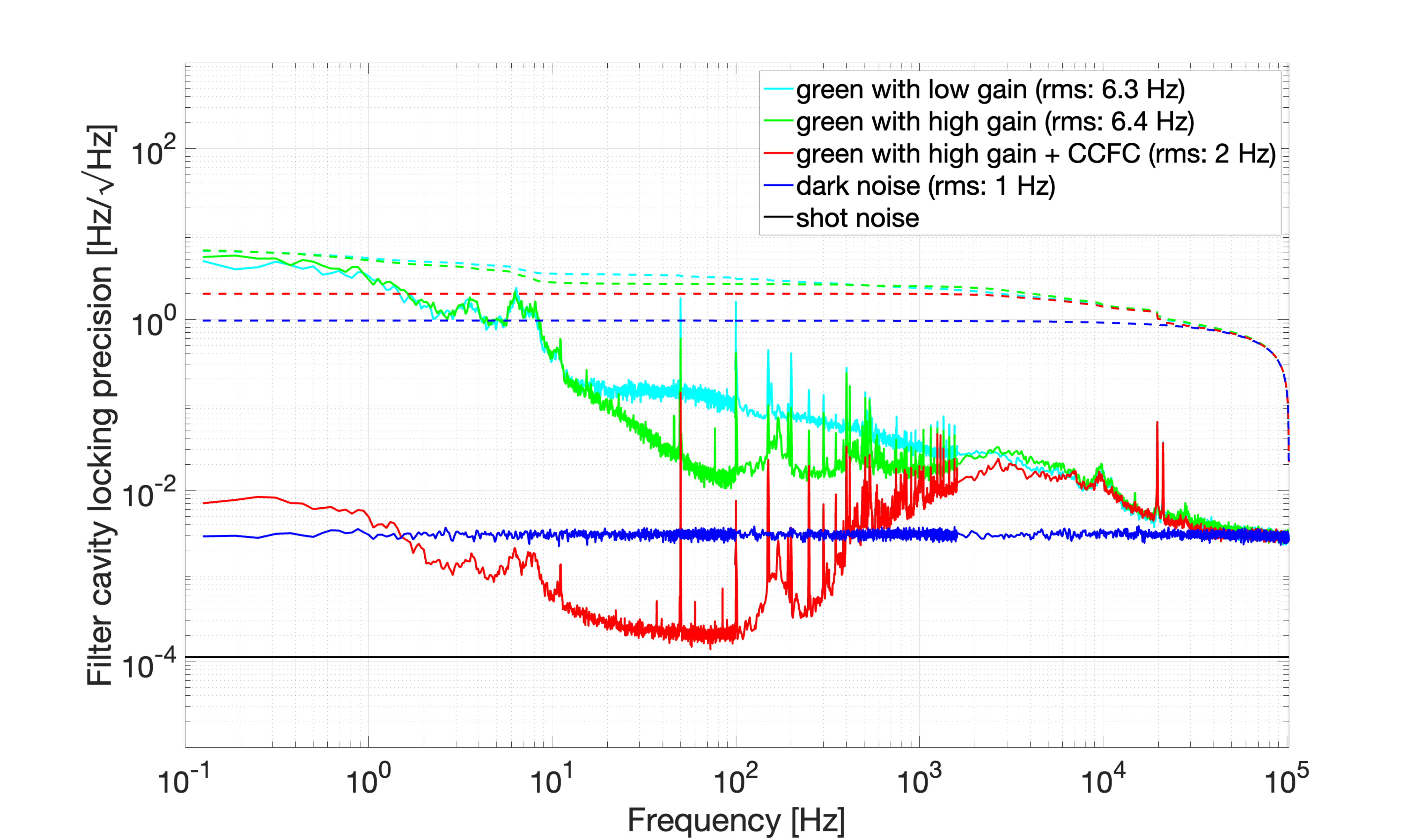}
\caption{Filter cavity locking precision with different filters of filter cavity control. The cyan and green curves are the measured spectra of the CCFC error signals when the filter cavity is controlled by the green field with low gain and high gain, respectively. The red curve is the measured spectrum of the CCFC error signal when the filter cavity is controlled by CCFC and the green field with high gain. The blue curve is the measured spectrum of the dark noise and the black line is the expected shot noise.}
\label{CCFC lock precision}
\end{center}
\end{figure}

\par The measured locking precision with different control filters is shown in Fig. \ref{CCFC lock precision}. The locking precision is limited by the filter cavity length noise below 10 Hz, the laser frequency noise between 10 Hz and 50 kHz, and the dark noise above 50 kHz. The dark noise mainly comes from the electronic noise of the photodetector. Figure \ref{CCFC lock precision} shows that by increasing the gain of the green control, the laser frequency noise above 10 Hz is reduced, while the filter cavity length noise below 10 Hz is not reduced. This is because the laser frequency noise is common for both the IR and green fields, while the filter cavity length noise is different for them. This is the reason why the CCFC is necessary to reduce the filter cavity length noise below 10 Hz.
\par By adding the CCFC, the locking precision (rms) is improved from 6.4 to 2 Hz (from 6.8 to 2.1 pm in the filter cavity length noise), as shown in Fig. \ref{CCFC lock precision}. The improvement of locking precision (rms) from 6.4 to 2 Hz comes from the improvement of the filter cavity length noise of 3.7 Hz and the frequency noise of 0.7 Hz. Although the improvement of the frequency noise can also be realized by the larger gain of the green control, the improvement of the filter cavity length noise can be realized only by CCFC. 
\par In Fig. \ref{CCFC lock precision} the locking precision with CCFC is even below the dark noise between 2 and 400 Hz, which means that the actual locking precision should be limited by the dark noise between 2 and 400 Hz. However, the rms of the dark noise below 400 Hz is smaller than 0.1 Hz and this will not affect the actual rms of the locking precision with CCFC.
\par The rms of the locking precision with CCFC is dominated by the laser frequency noise and the dark noise. In GW detectors, the squeezer laser (main laser in our experimental setup) will be locked to the interferometer laser and the filter cavity length control will be fed back to the filter cavity mirror only. As the interferometer laser is much more stable than the squeezer laser, the frequency noise of the squeezer laser is expected to be reduced in GW detectors compared to our experimental setup. The dark noise will also be reduced in GW detectors as the CCFC error signal can be obtained at the reflection of an output mode cleaner without introducing the pick-off beam splitter. Therefore, the rms of the locking precision with CCFC will be improved in GW detectors.

\subsection{Frequency-dependent squeezing with CCFC}
\begin{figure}[h] 
\begin{center}
\includegraphics[width=15cm,bb= 100 0 3070 1100]{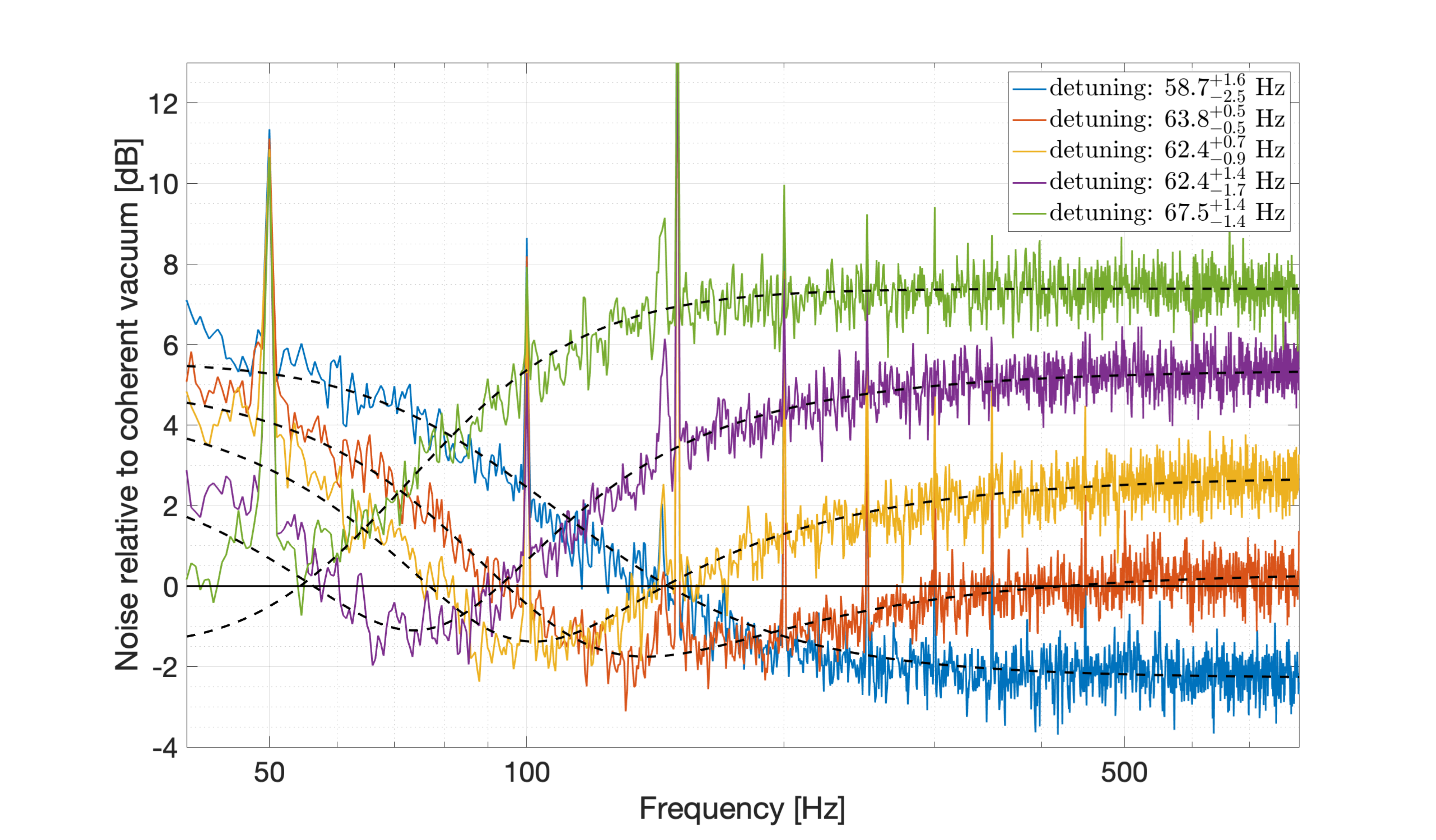}
\caption{Noise spectra of the frequency-dependent squeezing with CCFC for different homodyne angles. Each curve has been fitted, assuming the degradation parameters listed in Table \ref{CCFC degradation parameters}, to extract the homodyne angle, cavity detuning, propagation losses, and generated squeezing. Each spectrum has a resolution of 0.5 Hz and is averaged 100 times, leading to an acquisition time of 200 s.}
\label{CCFC FDS}
\end{center}
\end{figure}

The frequency-dependent squeezing with the CCFC is measured for different homodyne angles. The measured spectra using the CCFC are shown in Fig. \ref{CCFC FDS}. The measured spectra are fitted using Markov chain Monte Carlo methods. The fitting is started from 60 Hz because the backscattering noise becomes dominant below 60 Hz. The fitting parameters are the homodyne angle, cavity detuning, propagation losses, and generated squeezing. 
\par The average of the fitted cavity detuning in Fig. \ref{CCFC FDS} is about 63 Hz, which corresponds to the squeezing rotation frequency of about $63\  \mathrm{Hz}\times \sqrt{2} = 89$ Hz \cite{PhysRevD.90.062006}. The detuning fluctuation obtained from the measurement is about 9 Hz, and this value is better than that without the CCFC and the alignment control with the green field, which is $\sim 30$ Hz, as shown in Ref. \cite{PhysRevLett.124.171101}. 

\begin{table}[t]
 \begin{center}
 \caption{Squeezing degradation parameters with CCFC. The propagation losses and generated squeezing are obtained from the fitting result in Fig. \ref{CCFC FDS}, while the other parameters are measured values.}
 \label{CCFC degradation parameters}
  \begin{tabular}{c c}
  \hline  \hline
  Parameter & Current \\
  \hline 
 Propagation losses &  52 $\pm$ 3 \% \\ 
 Filter cavity losses & 120 $\pm$ 30 ppm \\
 Mode-mismatch squeezer-filter cavity &  6 $\pm$ 1 \% \\ 
Mode-mismatch squeezer-local oscillator &  2 $\pm$ 1 \%  \\ 
Frequency-independent phase noise (rms) & 30 $\pm$ 5 mrad \\ 
Filter cavity length noise (rms) &  2.0 $\pm$ 0.5 pm \\ 
Generated squeezing & 9.6 $\pm$ 0.6 dB \\
  \hline \hline
  \end{tabular}
  \end{center}
  \end{table}

\section{Discussion} 
In Fig. \ref{CCFC FDS} the detuning of the filter cavity fluctuates by 9 Hz with CCFC. This detuning fluctuation of 9 Hz includes the residual detuning fluctuation of 2 Hz in Fig. \ref{CCFC lock precision}, the detuning fluctuation caused by the fluctuation of the zero-crossing point of the CCFC error signal, and the possible detuning fluctuation caused by filter cavity control with green field. One of the reasons for the detuning fluctuation is the IR alignment fluctuation of the filter cavity, as explained in Sec. \ref{Principle}. To explain the detuning fluctuation of 9 Hz by only the IR alignment fluctuation, the fluctuation of the mode mismatch between the squeezer and the filter cavity caused by the IR alignment fluctuation needs to be 22 \%. However, as listed in Table \ref{CCFC degradation parameters}, the fluctuation of the mode mismatch between the squeezer and filter cavity is 2 \%, and it is difficult to explain the detuning fluctuation of 9 Hz with this mechanism. We leave the investigation of the reason for the detuning fluctuation of 9 Hz with CCFC to future work. For long-term operation of the filter cavity, the alignment control of the filter cavity with CCFC will be necessary to stabilize the detuning fluctuation caused by the IR alignment fluctuation.
\par In this experiment, we use a combination of CCFC and green error signals for filter cavity control. As the filter cavity length noise is different for the IR and green fields, the control with the green field should be removed after CCFC is engaged. We will control the filter cavity with only CCFC in the future.
\par
\begin{figure}[t] 
\begin{center}
\includegraphics[width=13.5cm,bb= 100 0 3100 1300]{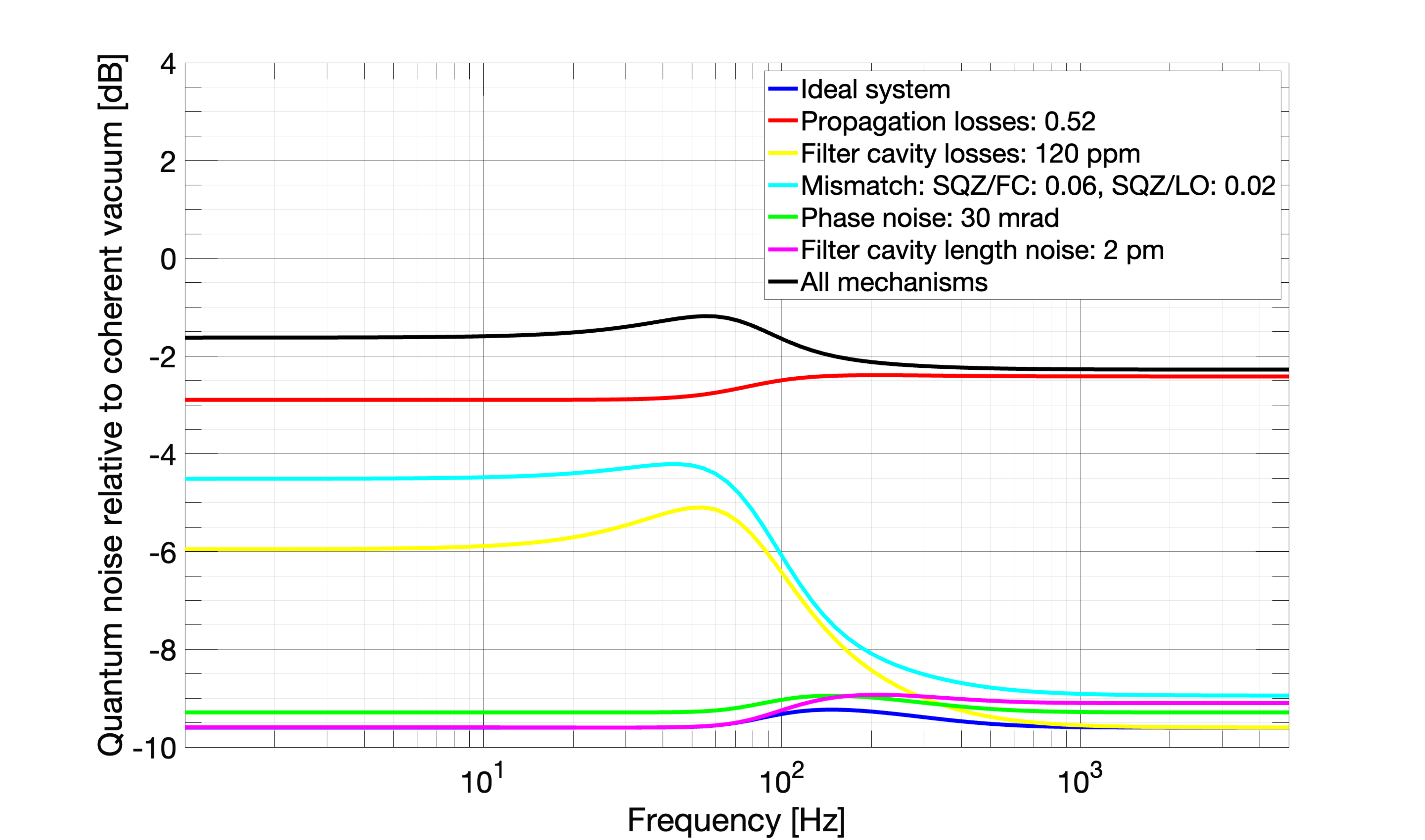}
\caption{Squeezing degradation budget with CCFC. The black curve shows the expected improvement of the quantum noise for a GW detector such as KAGRA. The squeezing degradation parameters used in this figure are listed in Table \ref{CCFC degradation parameters}. As explained in Ref. \cite{PhysRevD.102.042003}, the filter cavity length noise is suppressed at low frequencies with CCFC, while it is not at high frequencies.}
\label{CCFC degradation}
\end{center}
\end{figure}

 The squeezing degradation budget with CCFC is shown in Fig. \ref{CCFC degradation}. The quantum noise reduction is limited by propagation losses at all frequencies. The main loss sources are the pick-off beam splitter for CCFC, the OPO, and the in-vacuum Faraday isolator. The pick-off beam splitter introduces a loss of 20 \%, but it can be removed in real GW detectors because the CCFC error signal can be obtained at the reflection of an output mode cleaner without introducing the pick-off beam splitter. The OPO and in-vacuum Faraday isolator introduce losses of $\sim$ 10 \% and 15 \%, respectively. Both components will be replaced with low-loss ones. In particular, there are Faraday isolators with losses below 1 \% \cite{Genin:18} and OPOs with escape efficiencies as high as 99 \% \cite{PhysRevLett.117.110801}. By replacing these components, it is possible to achieve propagation losses of 10 \%. 
\par The mode mismatch between the squeezer and filter cavity is $\sim$ 6 \%, and this can be reduced by the alignment control of the filter cavity using CCFC.
\par With these improvements, it is possible to achieve a quantum noise reduction of 6 dB at high frequencies and 4 dB at low frequencies, and the sensitivity of GW detectors will be significantly increased.

\section{Conclusion}
For accurate length and alignment control of a filter cavity, we experimentally demonstrated the new control scheme for a filter cavity with coherent control sidebands. In addition to the conventional filter cavity control with the green field, we successfully controlled the length of a 300-m filter cavity with the coherent control sidebands. By adding the length control with the coherent control sidebands, the filter cavity length noise (rms) was reduced from 6.8 to 2.1 pm.

\begin{acknowledgments}
We thank the Advanced Technology Center (ATC) of NAOJ for its support. This work was supported by the JSPS Grant-in-Aid for Scientific Research (Grants Nos. 15H02095, 18H01235, and 21H04476), the LabEx UnivEarthS (ANR-10-LABX-0023 and ANR-18-IDEX-0001), and the EU Horizon 2020 Research and Innovation Program under the Marie Sklodowska-Curie Grant Agreement No. 734303  (NEWS) and 101003460 and GA 101003460 (PROBES).
N. A. was supported by the JSPS Grant-in-Aid for Scientific Research (Grant No. 18H01224), the JSPS Grant-in-Aid for Challenging Research (Exploratory) (Grant No. 18K18763), and the JST CREST (Grant No. JPMJCR1873). E. C., M. E., and M. P. were supported by the JSPS Grant-in-Aid for JSPS Fellows (Grants Nos. 18F18024, 20F20803, and 20F20713), respectively. H. L. and H. V. were supported by the Deutsche Forschungsgemeinschaft (DFG, German Research Foundation) under Germany's Excellence Strategy - EXC 2123 QuantumFrontiers - 390837967.
\end{acknowledgments}


\bibliography{CC_PRD2}

\end{document}